\documentclass[twocolumn,showpacs,preprintnumbers,amsmath,amssymb,prl]{revtex4}

\usepackage{graphicx}
\usepackage{dcolumn}
\usepackage{bm}
\usepackage{amssymb}

\begin{document}


\title{Matter Coupling to Strong Electromagnetic Fields in Two-Level Quantum Systems with Broken Inversion Symmetry}

\author{O.V. Kibis$^1$}\email{Oleg.Kibis@nstu.ru}\author{ G.Ya. Slepyan$^2$}
\author{S.A. Maksimenko$^2$} \author{A. Hoffmann$^3$}

\affiliation{$^1$Department of Applied and Theoretical Physics,
Novosibirsk State Technical University, Karl Marx Avenue 20,
630092 Novosibirsk, Russia
\\
$^2$Institute for Nuclear Problems, Belarus State University,
Bobruyskaya St. 11, 220050 Minsk, Belarus
\\
$^3$ Institut f\"ur Festk\"orperphysik, Technische Universit\"at
Berlin, Hardenbergstra{\ss}e 36, D-10623 Berlin, Germany}


\begin{abstract}
We demonstrate theoretically the parametric oscillator behavior of
a two-level quantum system with broken inversion symmetry exposed
to a strong electromagnetic field. A multitude of resonance
frequencies and additional harmonics in the scattered light
spectrum as well as altered  Rabi frequency are predicted to be
inherent to such systems. In particular,  dipole radiation at the
Rabi frequency appears to be possible. Since the Rabi frequency is
controlled by the strength of coupling electromagnetic field, the
effect can serve for the frequency-tuned parametric amplification
and generation of electromagnetic waves. Manifestation of the
effect is discussed for III-nitride quantum dots with strong
build-in electric field breaking the inversion symmetry. Terahertz
emission from arrays of such quantum dots is shown to be
experimentally observable.
\end{abstract}

\pacs{42.50.Hz, 47.20.Ky, 78.67.Hc, 85.60.Jb}

\maketitle

The resonant interaction of quantum systems with strong
electromagnetic field
\cite{Cohen-Tannoudji_b98,Scully_b01
} is permanently in the focus of interest,  both due to  high
methodological value of arising problems and their direct relation
to current applied projects, such  as new generation of
high-efficient lasers \cite{bimberg_b99}, laser cooling of atoms
\cite{Scully_b01}, development of basis for quantum information
processing
\cite{
Blais_04}, etc. One of bright manifestations of the strong
field-matter coupling is the Rabi effect \cite{Scully_b01}:
oscillations of the level population in a quantum system exposed
to a monochromatic electromagnetic wave. The simplest physical
model leading to harmonic Rabi oscillations is a two-level
symmetrical quantum system placed in a given classical single-mode
electromagnetic field \cite{Scully_b01}. Incorporation into this
simplest model additional physical factors results in many
nontrivial effects. For example, accounting for the quantum nature
of light leads to the concept of radiation-dressed atoms
\cite{Cohen-Tannoudji_b98} and the `collapse-revivals' phenomenon
in the population dynamics of a system exposed to coherent light.
Time-domain modulation of the field-matter coupling constant
\cite{Law_96},  local-field effects in nanostructures
\cite{Slepyan_prb04,Slepyan_prb07} and phonon-induced dephasing
\cite{forstner_03} provide new possibilities for the control of
the Rabi oscillations. Many interesting effects manifest
themselves in more complex systems, such as two coupled Rabi
oscillators \cite{Tssukanov_prb06},  Rabi oscillators based on
superconducting electrical circuits \cite{Blais_04}, and systems
where Rabi oscillations are strongly influenced by intraband
motion of quasi-particles \cite{Steiner_08}. In the given Letter
we investigate  the role in the Rabi effect of a new physical
factor --- violation of the  inversion symmetry. Inherent in many
quantum systems, the factor is  ignored by the conventional
physical model \cite{Scully_b01} of Rabi oscillations. Meanwhile,
our theoretical analysis predicts pronounced manifestation of the
violation in a set of observable effects in different physical
systems.

Let us consider a two-level quantum system  with $|a\rangle$ and
$|b\rangle$ as excited and ground states, respectively. Let the
system interacts with a classical linearly polarized monochromatic
electromagnetic field. The system is described by the Hamiltonian
$\hat{\cal H}=\hat{\cal H}_0+\hat{\cal H}_\mathrm{int}$. The
free-particle Hamiltonian, written in the basis of these two
states, is $\hat{\cal H}_0=\hbar\omega_0\hat{\sigma}_z/2$, where
$\hat{\sigma}_z$ is the Pauli matrix and $\omega_0$ is the
resonant frequency of the two-level system. The interaction
Hamiltonian is expressed in terms of the amplitude $\mathbf{E}$
and frequency $\omega$ of the driving field by $\hat{\cal
H}_\mathrm{int}=-\mathbf{E}\hat{{\mathbf{d}}}\cos(\omega t)$,
where $\hat{{\mathbf{d}}}$ is the electric dipole moment operator
and $\mathbf{d}_{ij}=\langle i|\hat{{\mathbf{d}}}|j\rangle$ are
its matrix elements.  The critical assumption, which distinguishes
the systems being considered from standard ones, is the violation
of the inversion symmetry. Since in that case the states
$|a\rangle$ and $|b\rangle$ do not possess a certain spatial
parity, the diagonal matrix elements of the dipole moment operator
prove to be nonequivalent, $\mathbf{d}_{aa}\neq \mathbf{d}_{bb}$,
dictating thus physical effects described hereafter.

We shall seek the solution of the Schr\"odinger equation with the
Hamiltonian described above in the form of
$|\psi\rangle=C_a(t)|a\rangle+C_b(t)|b\rangle$. Substituting this
expression into the Schr\"odinger equation, we arrive at the
equations
\begin{eqnarray}
i\hbar\dot{C}_a=\Bigl[\frac{\hbar\omega_0}{2}-\mathbf{E}\mathbf{d}_{aa}\cos(\omega
t)\Bigr]C_a- \mathbf{E}\mathbf{d}_{ab}\cos(\omega
t)C_b\,,\,\,\,\label{6a}
    \\
i\hbar\dot{C}_b=\Bigl[-\frac{\hbar\omega_0}{2}-\mathbf{E}\mathbf{d}_{bb}\cos(\omega
t)\Bigr]C_b-\mathbf{E}\mathbf{d}_{ba}\cos(\omega t)C_a\,.\label{6}
\end{eqnarray}
In order to solve these equations with respect to $C_a$ and $C_b$,
we first rewrite them for modified amplitudes
$c_{a,b}=C_{a,b}\exp[\pm i\omega_0t/2-i\phi_{a,b}(\omega,t)]$,
where $\phi_j(\omega,t)=\mathbf{E}\mathbf{d}_{jj}\sin(\omega
t)/\hbar\omega$, and signs $\pm$ correspond to indexes $a$ and
$b$, respectively. The system of equations \eqref{6a}--\eqref{6}
is then reduced to the form as follows
\begin{eqnarray}
&&\hbar\dot{c}_a=
    i\mathbf{E}^\ast_\mathrm{eff}\mathbf{d}_{ab}\cos(\omega t)c_b
    e^{i\omega_0t}\,,\label{5a}\\
\rule{0in}{2ex}
&&\hbar\dot{c}_b=i\mathbf{E}_\mathrm{eff}\mathbf{d}_{ba}\cos(\omega
t)c_a
    e^{-i\omega_0t}\,,\label{5}
\end{eqnarray}
where the effective electric field strength is
\begin{equation}\label{Eeff}
\mathbf{E}_\mathrm{eff}(t)=\mathbf{E}e^{-i\kappa\sin(\omega
t)}=\mathbf{E}\sum^{\infty}_{n=-\infty}J_n(\kappa)e^{in\omega
t}\,,
\end{equation}
$J_n(\kappa)$ is the Bessel function of the first kind, and
$\kappa=\mathbf{E}(\mathbf{d}_{bb}-\mathbf{d}_{aa})/\hbar\omega$
is the symmetry violation parameter. Let us stress that
Eqs.~\eqref{5a}--\eqref{5} are analogous to standard equations of
two-level system \cite{Scully_b01} with the only difference that
the driving field amplitude  $\mathbf{E}$ is replaced by the
effective amplitude \eqref{Eeff}. That the effective amplitude
$\mathbf{E}_\mathrm{eff}(t)$ is time-dependent causes the system
to become a parametric oscillator with full set of intrinsic
properties, unusual for standard Rabi oscillators. In particular,
a multitude of resonant frequencies
$\omega=\omega_0/n\,,~n=1,2,3,...$ appears in the system. Further
we shall assume that the driving field frequency $\omega$ is in
the vicinity of the frequency $\omega_0/m$ ($m$-th resonance) and
the conditions
$|\mathbf{E}\mathbf{d}_{ab}mJ_m(\kappa)/\kappa\hbar(\omega_0-m\omega)|\gtrsim1$
and
$|\mathbf{E}\mathbf{d}_{ab}nJ_n(\kappa)/\kappa\hbar(\omega_0-n\omega)|\ll1$
($n\neq m$) are fulfilled, allowing us to neglect interaction of
the two-level system with all harmonics \eqref{Eeff} other than
harmonics $m\pm1$. Then Eqs.~\eqref{5a}--\eqref{5} resume the form
analogous to equations for symmetric two-level systems and can
easily be solved by invoking the rotating-wave approximation
\cite{Scully_b01}. Assuming the nondiagonal dipole matrix elements
to be real-valued, we obtain
\begin{eqnarray}
&&C_a(t)=\left\{C_a(0)\left[\cos\Bigl(\frac{\Omega t}{2}\Bigr)-
\frac{i\Delta}{\Omega}\sin\Bigl(\frac{\Omega t}{2}\Bigr)\right]
\right.\nonumber\\
&&\qquad\left.+i\frac{\Omega_R}{\Omega}C_b(0)\sin\Bigl(\frac{\Omega
t}{2}\Bigr)\right\}e^{-im\omega t/2}e^{i\phi_a(\omega,t)}\,, \label{7a}\\
&&C_b(t)=\left\{C_b(0)\left[\cos\Bigl(\frac{\Omega t}{2}\Bigr)
+\frac{i\Delta}{\Omega}\sin\Bigl(\frac{\Omega t}{2}\Bigr)\right]
\right.\nonumber\\
&&\qquad\left.+i\frac{\Omega_R}{\Omega}C_a(0)\sin\Bigl(\frac{\Omega
t}{2}\Bigr)\right\}e^{im\omega
t/2}e^{i\phi_b(\omega,t)}\,,\label{7}
\end{eqnarray}
where the parameters $\Delta=\omega_0-m\omega$,
$\Omega=\sqrt{\Omega_R^2+\Delta^2}$, and the Rabi frequency
\begin{equation}
\Omega_R=2\mathbf{E}\mathbf{d}_{ab}mJ_m(\kappa)/\kappa\hbar\label{Rabi}
\end{equation}
are written in the vicinity of $m$-th resonance. Since we consider
the strong coupling regime, the driving field $\mathbf{E}$ is
assumed to be sufficiently strong, $\Omega_R\tau\gg1$, to neglect
impact of the linewidth $\hbar/\tau$.  Note that the Rabi
frequency in systems with broken inversion symmetry \eqref{Rabi}
shows nonmonotonic dependence on the driving field strength and
even can turn zero at those values of $\mathbf{E}$ which
correspond to roots of the Bessel function $J_m(\kappa)$.

It is well known that in symmetrical two-level systems the Rabi
effect manifests itself in the power spectrum of the scattered
light by peaks centered at the incident light frequency $\omega$
and at the displaced frequencies $\omega\pm\Omega$ (Mollow
triplet, \cite{Mollow_69}). In order to reveal peculiarities of
the scattered light induced by the symmetry violation, it is
suffice to consider the process in the framework of classical
electrodynamics. It allows analyzing electronic subsystem, in
respect to irradiation of electromagnetic waves, as a classical
dipole with the oscillating dipole moment $
\mathbf{d}(t)=\langle\psi|\hat{{\mathbf{d}}}|\psi\rangle$. For
definiteness, we identify  initial time as that corresponding to
system being in the excited state, i.e., $C_a(0)=1$ and
$C_b(0)=0$. Then, substituting wave function $|\psi\rangle$ with
coefficients \eqref{7a}--\eqref{7} into expression for the dipole
moment and omitting the time-independent terms, we arrive at the
expression
\begin{eqnarray}\label{12}
&&\mathbf{d}(t)=(\mathbf{d}_{aa}-\mathbf{d}_{bb})\frac{\Omega_R^2}{4\Omega^2}e^{i\Omega
t}-\mathbf{d}_{ab}\frac{\Omega_R}{2\Omega}
    \nonumber\\
    &&~~
\times\sum_{n=-\infty}^{\infty}J_{m-n}(\kappa) e^{in\omega
t}\left[\frac{\Delta}{\Omega}
+\frac{1}{2}\left(1-\frac{\Delta}{\Omega}\right)e^{-i\Omega
t}\right.
    \nonumber\\
    &&~~
-\left.\frac{1}{2}\left(1+\frac{\Delta}{\Omega}\right) e^{i\Omega
t}\right]+\mathrm{c.c.}
\end{eqnarray}
for the dipole moment in the vicinity of $m$-th resonance. As
follows from Eq.~\eqref{12}, the radiation spectrum consists of a
singlet at the frequency $\Omega$, and an infinite sequence of
triplets with the frequencies $n\omega,\,n\omega\pm\Omega$ (see
Fig.~1). It should be emphasized that the frequency multiplication
is the general property of systems with broken inversion symmetry
and can be used for the high harmonic generation
\cite{Calderon99}.
\begin{figure}[th]
\includegraphics[width=0.48 \textwidth]{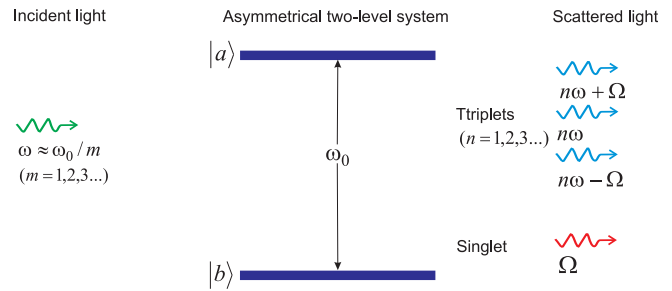}
\caption{Schematics of the light scattering in a two-level system
with broken inversion symmetry in the vicinity of one of the
possible resonances $\omega=\omega_0/m$.}\label{fig1}
\end{figure}
Amplitudes of harmonics of the triplets rapidly decrease with
increasing $n$ or $m$, while the singlet $\Omega$ amplitude
depends on neither $n$ nor $m$. Radiation of the dipole \eqref{12}
in the vicinity of each resonance is characterized by its own Rabi
frequency \eqref{Rabi} decreasing with the resonance number $m$
increase. In systems with inversion symmetry diagonal elements of
the dipole moment are identical. As a result, higher resonances
($m>1$), higher triplets ($n>1$), and the singlet vanish. In that
case, coefficients \eqref{7a}--\eqref{7} and expression for the
dipole moment \eqref{12} coincide with the solution of the problem
presented in Ref.~\cite{Scully_b01}.

Further we confine consideration to systems with a weak violation
of the inversion symmetry, when the condition $|\kappa|\ll1$ holds
true. In such systems the asymmetry effect is expected to be most
pronounced in the vicinity of the first resonance $m=1$
$(\omega\approx \omega_0)$. If then we restrict ourselves to the
most interesting case of resonant driving field, when $\Delta=0$
and $\Omega=\Omega_R$, we reduce  \eqref{12} to the expression
\begin{equation}\label{13}
\mathbf{d}(t)=\frac{\mathbf{d}_{aa}-\mathbf{d}_{bb}}{2}\cos(\Omega_Rt)-
\mathbf{d}_{ab}\sin(\omega_0t)\sin(\Omega_Rt)\,,
\end{equation}
where Rabi frequency \eqref{Rabi} takes its conventional form
$\Omega_R=\mathbf{E}\mathbf{d}_{ab}/\hbar$. In typical quantum
systems the Rabi frequency is much less then the driving field
frequency $\omega=\omega_0$. Therefore, in addition to the
high-frequency harmonics $\omega_0\pm\Omega_R$ (second term in the
right-hand part of Eq. \eqref{13}), violation of the inversion
symmetry leads to the irradiation of low-frequency electromagnetic
waves at Rabi frequency (first term in the right-hand part of Eq.
\eqref{13}) with the time-averaged radiation intensity
\begin{equation}
I_R=\frac{|\mathbf{d}_{aa}-\mathbf{d}_{bb}|^2}{12c^3}\Omega_R^4.\label{14}
\end{equation}
The quantity $|\mathbf{d}_{aa}-\mathbf{d}_{bb}|$ is further
referred to as effective dipole moment. Note that the frequency of
the scattered radiation $\Omega_R$ depends only on the driving
field $\mathbf{E}$ and does not depend on both the frequency of
the scattering system $\omega_0$ and the frequency of incident
light $\omega$.

The simplest quantum system devoid of the inversion center is a
Rydberg hydrogen atom imposed to a homogeneous static electric
field  ${\cal E}$. Assuming the atom to be in ground state, we
find that the Rabi frequency is expressed in terms of the Bohr
radius $a_B=\hbar^2/m_ee^2$ by
$\Omega_{R}=4\left({2}/{3}\right)^5e\,a_BE/\hbar$.
Correspondingly, for the electric field ${\cal E}$  much less than
the intra-atomic electric field $e/a_B^2$, the intensity of the
radiation at the Rabi frequency is determined by Eq.~\eqref{14}
with the effective dipole moment given by
$|\mathbf{d}_{aa}-\mathbf{d}_{bb}|=(1/8)(4/3)^{11}\, {a_B^3\cal
E}$. This expression  is applicable not only to hydrogen atom but
also can be used for estimating parameters of the emission at the
Rabi frequency in arbitrary quantum systems with broken inversion
symmetry. For that aim, the Bohr radius $a_B$ should be replaced
by the characteristic linear extension of the system. Therefore,
the radiation intensity \eqref{14} rises with the increase of the
system size. In that connection, confined semiconductor
nanostructures with discrete energy spectrum and linear extension
multiply exceeding the atomic size, quantum dots (QDs)
\cite{bimberg_b99}, can serve as prospective systems for the
effect observation.

In quantum dots, driving electromagnetic field transfers electrons
from valence band into conduction band; what is why in QDs ground
state $|b\rangle$ corresponds to the absence of free carriers
while first excited state $|a\rangle$ is the state with electron
in the conduction band and hole in the valence band. In that case,
$\hbar\omega_0$ is approximately equal to the semiconductor
bandgap, and the Rabi frequency is determined by the standard
expression $\Omega_R=\mathbf{E}\mathbf{d}_{cv}/\hbar$ with the
dipole matrix element $\mathbf{d}_{cv}$ corresponding to interband
transitions. Among a variety of different types of QDs,
nitride-based confined structures seem to be most promising for
the effect observation (for  parameters of III-nitrides AlN, GaN,
and InN see Ref. \cite{Rinke_08}). Indeed, GaN and similar
III-group nitride alloys have a hexagonal (wurtzite) structure. As
a consequence of giant piezoelectric effect inherent in hexagonal
crystals, QDs based on structures AlN/GaN and InN/GaN have a
strong built-in strain-induced electric field with strength
$\cal{E}$ of several MV/cm \cite{Widmann_98,Moriwaki_00}. Due to
the strong electric field, conduction-band electrons and
valence-band holes in the QDs get spatially separated
\cite{Williams_05,Bretagnon_06}, and the effective dipole moment
of a III-nitride QD is given by simple relation
$|\mathbf{d}_{aa}-\mathbf{d}_{bb}|\sim el$, where $l$ is the QD
height. Then, for typical $l$ of several nanometers, the effective
dipole moment of nitride QDs is estimated as
$|\mathbf{d}_{aa}-\mathbf{d}_{bb}|\sim 10\,$Debye, what is tens
thousands times as large as the effective dipole moment of
hydrogen atom in the same electric field.

Since the Rabi frequency depends on the driving field strength
$\mathbf{E}$, the broken inversion symmetry-induced low-frequency
singlet $\Omega_R$ in the dipole emission spectrum can be used for
the tunable generation of electromagnetic waves at Rabi frequency
with intensity \eqref{14}. This problem is especially challenging
for frequency ranges where traditional methods either fail or
inefficient, such as terahertz domain.  Since this domain lies
between radio and optical frequency ranges, neither optical nor
microwave techniques are directly applicable for generating THz
waves. Therefore, a search for effective THz radiation sources is
one of most excited fields of modern applied physics
\cite{Ferguson_02,Dragoman_04,Lee_07}. One of the latest trends to
fill the THz gap is using nanostructures as THz emitters and
detectors. The quantum cascade THz transitions in QD systems
\cite{Gmachl_01,Chakraborty_03} and different electron mechanisms
of THz emission from carbon nanotubes
\cite{Kibis_05,Kibis_07_NL,Nemilentsau_07_PRL} have been proposed
and are being actively studied. Thus, the proposed mechanism of
THz emission fits well the current tendencies in nanophotonics.

Parameters of the THz emission from III-nitride QDs can be
evaluated in the following way. Using the estimate ${d}_{cv}\sim
10\,$Debye for the QD interband dipole moment \cite{Rinke_08}, we
obtain that the Rabi frequency turns out to be lying in the THz
range at relatively weak strength of the driving field,
$E\sim10^{5}\,$V/cm. Then, substituting the previously obtained
value of the effective dipole moment
$\left|\mathbf{d}_{aa}-\mathbf{d}_{bb}\right|\sim10\,$Debye into
Eq.~\eqref{14}, for the typical QD area $\sim10^{-12}$ cm$^{2}$ we
arrive at the radiative intensity per single QD $\sim10^{-11}$
W/cm$^2$ in the THz range. This estimate multiples as $N^2$ for
array of $N$ identical QDs with the lateral extension less than
the Rabi wavelength $\lambda_R=2\pi c/\Omega_R$. In such an array,
all QDs emit waves in phase. The THz-range wavelength
$\lambda_R\sim 10^{-2}$ cm restricts the array lateral size.
Taking into account that the typical density of nitride QDs is
$\sim 10^{11}$~cm$^{-2}$, we can estimate the number of QDs in
such an array by $N\sim10^7-10^8$. Therefore, the THz emission
power from the submillimeter-sized QD array may approach the
micro-Watt level, which is characteristic for the
resonant-tunneling diodes based on carbon nanotubes
\cite{Dragoman_04_1}. Certainly, state-of-the-art THz quantum
cascade lasers demonstrate essentially larger output power
\cite{Williams_07}. However, array of QDs with broken inversion
symmetry provides smooth frequency tuning by the variation of the
driving field intensity. The reducing impact of inhomogeneous
broadening is beyond of the estimate and will be considered
elsewhere. As to the homogeneous mechanisms of the line
broadening, like phonon scattering, they are accounted for
phenomenologically in the linewidth $\hbar/\tau$ and do not impose
any specific restrictions on the observability of the effect, but
just influence the criterion of the strong light-matter coupling
regime. Thus, the upper estimate presented of the radiation output
allows proposing the broken inversion symmetry-induced mechanism
of the radiation for the development of novel-type THz emitters
based on QDs avoid inversion symmetry. Obviously, parametric
amplification of THz radiation in  QD arrays is also possible and
can be applied for THz detecting.

Once more type of asymmetrical two-level quantum systems where the
discussed effect can be observed is superconducting quantum qubits
\cite{Blais_04,Aravantinos_05}. Away from the charge degeneracy
point of the superconducting quantum circuits \cite{Blais_04}
formed by Josephson qubits in microstrip resonators, the qubit
symmetry is broken and the Hamiltonian of the system (see Eq.~(16)
in Ref.~\cite{Blais_04}) takes the form analogous to that used in
our analysis. At that, the resonant transition frequency amounts
to $\sim 10\,$GHz while the Rabi frequency lies in the range $\sim
100\,$MHz. Intensity of low-frequency line in the spectrum of Rabi
oscillations can be controlled by the changing of dc gate voltage.
Chiral nanostructures, including chiral carbon nanotubes
\cite{Kibis_02_PE}, should also be noted as prospective systems
for observation of the effect, since the chirality breaks the
inversion symmetry. New interesting effects are expected for
asymmetric two-level oscillators placed inside band-gap structures
like photon crystal, microcavity or nanoantenna. For example, by
corresponding manipulation of the photonic crystal parameters, the
frequencies $\omega$ and $\omega_0$ can be chosen laying in the
band gap while the frequency $\Omega$ lies in the transparency
band. In that case, the strong coupling regime is realized for the
pump field, while at the frequency $\Omega$ the photonic crystal
serves as an antenna transforming near field into far field.
Similar effect appears in microcavity with resonant frequency
close to $\omega$ and leaking modes at the frequency $\Omega$.

In the paper we have assumed that the incident light pulse is much
longer than periods of both incident and scattered electromagnetic
field. When a collection of oscillators with broken inversion
symmetry is illuminated by an extremely short pulse, the
difference between amplitudes  $E$ and $E_{\rm{eff}}$ in
Maxwell-Bloch equations may result in the failure of the area
theorem and in the related effect of the carrier-wave Rabi
flopping. Such an effect is observed in semiconductors when the
Rabi frequency becomes comparable to the band gap frequency
\cite{Mucke_prl_01}.

To conclude, let us stress  that violation of the inversion
symmetry is common property of quantum oscillators of different
physical origination and, consequently, the effect predicted is
expected to manifest itself in spectral characteristics of
different optical processes, e.g. resonant fluorescence of
molecules.

The research was partially supported by the INTAS project
05-1000008-7801, IB BMBF (Germany) project BLR 08/001, the EU FP7
TerACaN project FP7-230778, RFBR (Russia) project 08-02-90004,
`Development of Scientific Potential of Russian Higher Education'
Program (project 2.1.2/2115), and BRFFR (Belarus)  project
F08R-009. The work of O.V.K. and S.A.M. was partially carried out
during the stay at the Institut f\"{u}r Festk\"{o}rperphysik, TU
Berlin, and supported by DAAD and DFG, respectively.

\end{document}